\documentclass[11pt]{article}

\usepackage{graphicx} % Required for inserting images
\usepackage{biblatex} % Required for references
\usepackage{amsmath,amssymb} % Required for math and symbols

\usepackage{geometry}
\usepackage{hyperref}
\usepackage{multirow}
\usepackage{enumitem}

\usepackage{soul}
\usepackage{color}

\newcommand{\hlc}[2][yellow]{ {\sethlcolor{#1} \hl{#2}} }

\usepackage[usenames,dvipsnames]{xcolor}

\definecolor{dgreen}{RGB}{0,150,40}

\begin{document}

\begin{center}
{\Large \bf Maximizing science return by coordinating the survey strategies of Roman with Rubin, and other major facilities}
\end{center}

\begin{tabular}{ll}
 {\bf Roman Core Community Survey}    &  Galactic Bulge Time Domain Survey\\
 {\bf Scientific categories }         & Exoplanets and exoplanet formation \\
                                      & Stellar physics and stellar types \\
                                      & Stellar populations \\
 {\bf Additional scientific keywords} & \\
 {\bf Submitting author}              & R.A.~Street, Las Cumbres Observatory, \\
                                      & rstreet@lco.global\\
\end{tabular}\\
{\bf Contributing authors: } S.~Gough-Kelly, Jeremiah Horrocks Institute, UCLan, C.~Lam, UC Berkeley, A.~Varela, M.~Makler, International Center for Advanced Studies, Argentina, E.~Bachelet, IPAC, J.R.~Lu, N.~Abrams, A.~Pusack, S.~Terry, UC Berkeley, R.~Di~Stefano, CfA Harvard, Y.~Tsapras, M.P.G.~Hundertmark, Uni Heidelberg, R.J.J.~Grand, Astrophysics Research Institute, LJMU, T.~Daylan, Princeton, J.~Sobeck, University of Washington

\section*{Abstract}
The Nancy Grace Roman Space Telescope will be one of several flagship survey facilities operating over the next decade starting $\sim$2025.  The deep near-IR imaging that Roman will deliver will be highly complementary to the capabilities of other survey telescopes that will operate contemporaneously, particularly those that can provide data at different wavelengths and messengers, or different time intervals.  Combining data from multiple facilities can provide important astrophysical insights, provided the data acquisition is carefully scheduled, and careful plans are made for appropriate joint data analyses.  In this White Paper, we discuss the broad range of science that would be enabled by coordinating Roman observations of the Galactic Bulge with those of the Vera C. Rubin Observatory.  Specifically, we discuss how Roman's characterization of lensing events caused by exoplanets, stellar systems and stellar remnants can be enhanced by data from Rubin.  The same data will also be highly advantageous for the determination of stellar properties, and for distinguishing exoplanetary transits.  It will enable more accurate period-color-luminosity relationships to be measured for RR~Lyrae throughout the Milky Way Bulge and Bar, probing galactic structure and dynamics. But we stress that this is only a sample of the full potential and advocate for a more complete study to be made as a joint effort between these major projects.  In summary, we recommend:
\begin{enumerate}[leftmargin=*]
\setlength\itemsep{0em}
\item Close coordination between the groups responsible for survey strategy for the Roman Mission and Rubin Observatory to maximize the scientific return of the combined data products.  
\item Wherever possible, coordinating the timing of the Roman Bulge observation seasons with the long-baseline observations of Rubin.  In particular, if Rubin undertakes rolling cadence seasons, these would be most beneficial if they occurred during the gaps in Roman Bulge seasons.  We also highlight the value of acquiring contemporaneous observations in constraining the masses of free-floating planet microlensing events. 
\item A broader study of the scientific potential of coordinated scheduling and data analysis from major surveys should be funded, including science by all of the Roman Core Surveys, and considering a wide range of complementary facilities and catalogs.  
\item The development of metrics designed to evaluate how changes in the strategy of one survey impact the science return of another.  These should be used by both surveys.  
\end{enumerate}
We note that we do not suggest any changes beyond the established Science Requirements for the RGBTDS, in terms of survey footprint or filter selection.  
\pagebreak

% Suggested length for white papers is 2-3 pages of text, plus figures, tables and references.
\section{Scientific Motivations}
The combination of near-IR timeseries imaging from the Nancy Grace Roman Space Telescope with multi-band optical imaging from the Vera C. Rubin Observatory's Legacy Survey of Space and Time (LSST, see Appendix A), will be highly beneficial for numerous science goals from both flagship missions.  Here, we highlight the scientific benefits from the Roman Galactic Bulge Time Domain Survey (RGBTDS, see White Paper by Gaudi et al.).  

\begin{figure}
\begin{centering}
\includegraphics[width=1.0\textwidth]{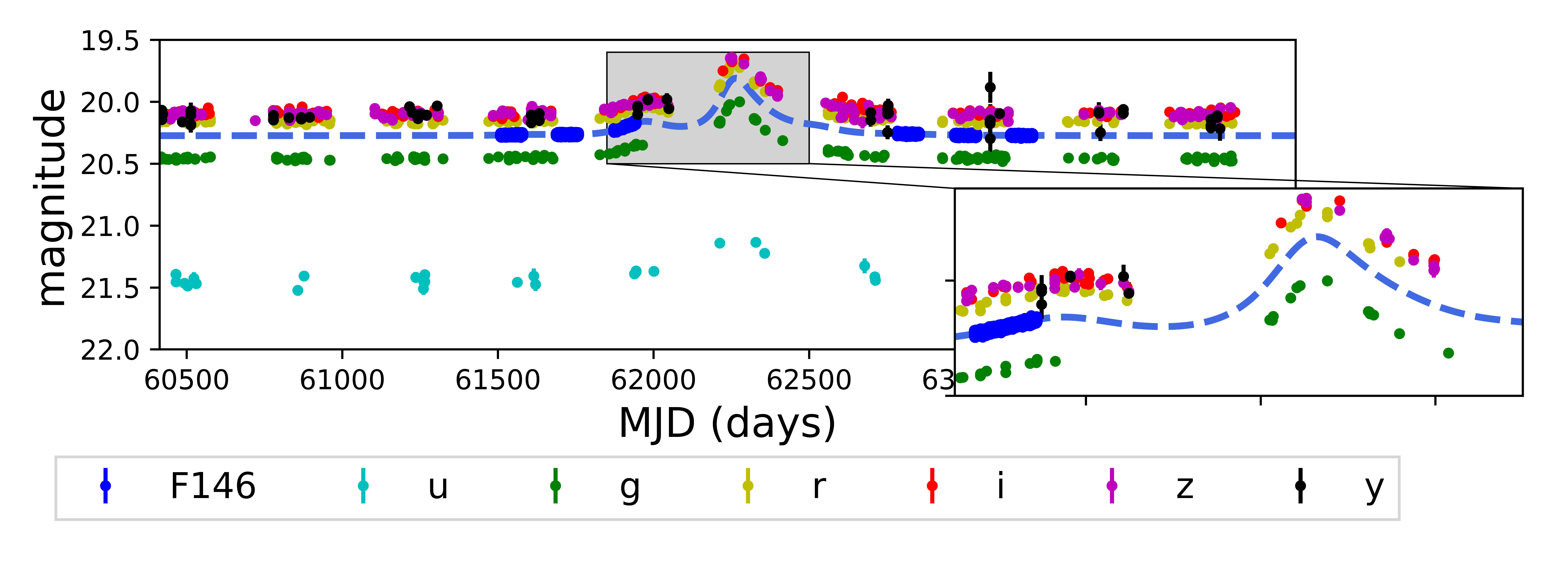} 
\caption{Simulation of the multi-wavelength lightcurves of a binary microlensing event, observed by both Roman and Rubin. The dotted line represents the event lightcurve observed in the Roman passband, while the lightcurves in the Rubin passbands are offset in magnitude due to the apparent magnitude of the source at these wavelengths (accounting for extinction). \label{fig:event_lightcurve}}
\end{centering}
\end{figure}

\noindent{\bf Microlensing: Characterizing exoplanets, brown dwarfs, and isolated compact objects}\\
The primary goal of the Roman GBTDS is to detect and characterize thousands of exoplanets in relatively distant ($\sim$1--10\,AU) orbits around their host stars, as well as free-floating planets \cite{Penny2019, Johnson2020}.  Both categories of planets can be detected from the gravitational lensing caused by their gravity as they pass in front of a background star \cite{Paczynski1986}.  RGBTDS is expected to characterize thousands of these cool-orbit and free-floating planets, thereby populating the last major gap in the distribution of known planets.  Owing to pointing constraints, Roman will observe the Bulge intensively (imaging every $\sim$15\,min) in `seasons' approximately 60--72\,d long, separated by gaps of several months (e.g., Fig.~\ref{fig:event_lightcurve}).  While Roman's exquisitely precise photometry and astrometry are expected to characterize most of the systems discovered, microlensing events are true transients, and the anomalous deviations (small, sharp bumps in the light curve) that betray the presence of planets can occur at any time during events which may last from 1 to over 100\,d.  Some of these anomalous signatures will be missed during the RGBTDS season gaps.  In addition, in order to characterize an event, it is crucial to measure the microlensing parallax \cite{Gould2000}. This is primarily achieved through {\bf regular observations throughout the event, including of the rising and falling `wings' of the lightcurve}, so consistent monitoring over a long time baseline is essential.  Furthermore, many short events will occur entirely within the gaps in the RGBTDS.  In these cases, the RGBTDS photometry at baseline, and timeseries astrometry will still be important to characterize the host star of the event, thereby allowing the parameters of the planetary systems to be measured. One of the mantras of contemporary exoplanet studies is "To know the planet, know the star". The same holds true for exoplanetary microlensing events where planetary masses and orbital properties require observations and modeling of both the planetary microlensing anomaly and the host star's longer-duration microlensing event light curve. Given that most stellar microlensing timescales are $t_{\mathrm{E}}\geq20$\,days and that $>4\times t_{\rm{E}}$ must be monitored to characterize the host star, the Roman 70-day observing windows followed by large gaps are simply too short. Yet few observatories can survey a large area to the limiting magnitudes reached by Roman.  
Rubin has a vital role to play in filling in the gaps and maximizing Roman's scientific potential.  

RGBTDS will not only explore a new population of exoplanets, but also thousands of stellar, brown dwarf, and compact object systems as well.  Microlensing is uniquely sensitive to even isolated Black Holes \cite{Sajadian2023} (see White Paper by Lam et al.{\em ``Characterizing the Galactic population of isolated
black holes''}).  Consisting of the products of massive star evolution, and perhaps also Primordial Black Holes \cite{Pruett2022}, determining the galactic population of Black Holes is a key constraint on theories of star formation and evolution \cite{Wiktorowicz2019}.  Black Holes can be detected as components of {\em binary systems}, for example from radiation during accretion of material from a companion \cite{Tsuna2019}, or from the Gravitational Wave signal produced during a merger \cite{Abbott2016}.  But the presence of a binary companion influences the evolution of an object, so determining the population of {\em isolated} Black Holes remains a key goal in understanding their development.  However, since the microlensing timescale is proportional to the square root of the lens mass, black hole  events are on average $100\times$ longer than planetary lensing events. As a result, {\bf no black hole lensing events will be fully observed by Roman} because of the seasonal gaps in Bulge visibility.  Very few other facilities are capable of surveying the full RGBTDS footprint to Roman's limiting magnitude, or of providing the crucial, long-baseline monitoring necessary to properly characterize these events.  Rubin offers a rare combination of a wide-field imager on a large aperture telescope, with a spatial resolution of 0.2"/pixel.  

\noindent{\bf Transiting Exoplanets:} The characterization of exoplanet host stars is particularly important.  For example, the stellar radius is an essential parameter required to determine the radii of the $\sim$60,000--200,000 transiting planets that RGBTDS is expected to discover \cite{Wilson2023, Tamburo2023}.  Metallicity information is also an important factor in models of planet formation: more massive planets tend to form around metal-rich stars \cite{Narang2018}, although the trend reverses for planets $>4\,M_{Jupiter}$.  As the Roman survey will explore planet frequency in previously unexplored stellar populations in the Galactic Disk and Bulge, fully characterizing the host stars will provide essential insights.  While the filter set for Roman's WFI does include some optical passbands (e.g. F062 480--760\,nm, roughly overlapping with SDSS-r), Rubin's ugrizy photometry will provide a critical extension to the wavelength range sampled.  A joint analysis of data from both missions will include metal lines that are key to accurately determining metallicities for the majority of stars, such as Ca~H\&K (393.3\,nm \& 396.9\,nm), Mg~I (515\,nm) and Na~D (580\,nm).  Data in SDSS-g, J, and K bands are recommended to reliably measure the metallicities even for M-dwarf stars, to $\pm\sim$0.08\,dex \cite{Hejazi2015}. 

\noindent{\bf Stellar Astrophysics and Kinematics:}
Multiwavelength imaging data will be important in determining stellar spectral types in the RGBTDS region as most sources will be too faint for spectroscopy.  Photometric metallicities derived from combined optical$+$NIR bandpasses will enable us to distinguish between metal-rich and -poor stellar populations that follow different rotation curves through the Galactic Center \cite{Clarkson2018} and which theory suggests support different Galactic structures \cite{GoughKelly22}.  Age estimates determined from isochrone fitting will be a key result, as Roman will reach the Main Sequence Turn Off at low Galactic latitude.  

\noindent{\bf Stellar Variability: } The timeseries data provided by Roman and Rubin will uncover a wealth of stellar variables, from eclipsing binaries to pulsators of all types.  Multi-band timeseries photometry will reveal the varying depth of stellar-companion eclipses in different passbands, constraining the spectral type of both companions and enabling them to be distinguished from planetary transits.  Time-variable color, particularly in passbands that are widely separated in wavelength, will be a vital parameter in the accurate classification of variable stars, since both Roman and Rubin will probe fainter limiting magnitudes than previous catalogs and complement Gaia \cite{Ivezic2015}.  
Color is also one of the essential terms in the period-color-luminosity relationship that allows RR~Lyrae to be used as standard candles.  A joint analysis of Roman and Rubin data would be highly beneficial to studies of the galactic structure in the ``heart of the Milky Way''.   
Roman GBTDS will be ideal for finding large populations of Long-Period Variables (LPVs) such as Miras and Semi-Regular Variables close to the Galactic Center, and its high cadence timeseries is ideal for asteroseismology.  Its deep limiting magnitude timeseries data will detect RR~Lyrae and LPVs deep within the Galactic Bulge and Bar and even on the far side, enabling us to map the 3D structure of the inner Milky Way, including extinction.   Thanks to Roman's precise parallax measurements, these stars will be used as standard candles to underpin cosmological models, and probe dust distributions in this region.  But the gaps in Roman's timeseries are problematic, as LPVs commonly have periods of 100s of days.  Rubin's photometry in the season gaps will constrain the morphology of the variable lightcurves and hence improve the measurement of their periodicities.  

\begin{figure}
\begin{centering}
\begin{tabular}{cc}
\includegraphics[width=0.45\textwidth]{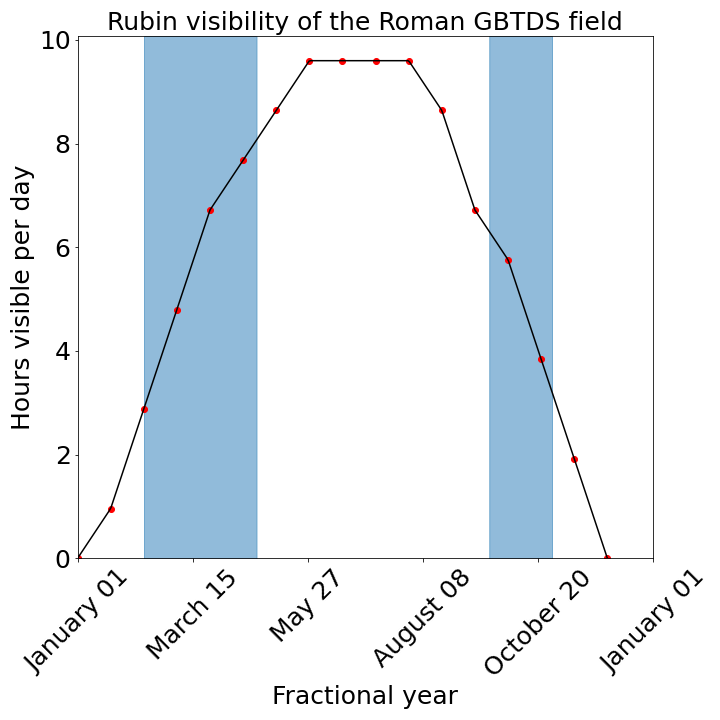} 
\includegraphics[width=0.45\textwidth]{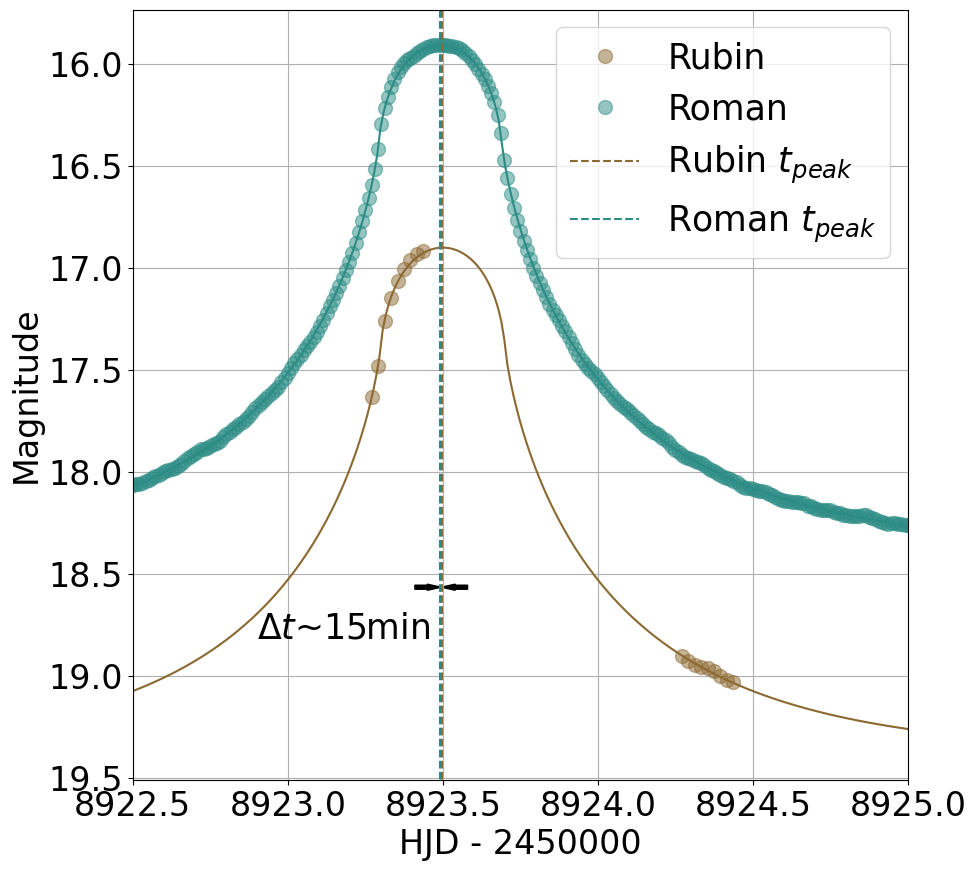} 
\end{tabular}
\caption{(Left) Illustration of the periods when simultaneous observations will be possible from both Roman and Rubin, assuming nominal dates for the RGBTDS seasons, indicated by blue shading. (Right) Simulation of a microlensing event lightcurve caused by a free-floating planet, observed by both Roman and Rubin during periods where contemporaneous observations are possible. Virtical dotted lines indicate the time of the event peak as measured form the two surveys.  The measureable time offset is a result of different lines of sight to the lensing event from the different observing platforms. 
\label{fig:bulge_visibility}}
\end{centering}
\end{figure}

\noindent{\bf Simultaneous Observations: } Although the nominal dates of the Roman GBTDS seasons (Feb--April, Sept--Oct) are not ideally timed for Earth-based observations, there is some overlap with the Bulge's annual visibility from Rubin's site in Chile, thanks to the telescope's 20$^{\circ}$ altitude pointing limit (Fig.~\ref{fig:bulge_visibility}).  This means that contemporaneous optical+NIR observations are possible for all stars in the field.  As noted in \cite{Street2018}, some of the Free Floating Planets detected by Roman during these overlap periods may also be detected by Rubin, allowing the microlensing satellite parallax of the event to be measured \cite{Yee2013}, and hence the mass of the lens.  While Roman will discover and characterize the morphology of the lightcurves of these elusive events, it cannot measure the masses of these lenses without this kind of additional constraint on the lensing parameters \cite{Johnson2020}.  
Such contemporaneous observations would also benefit stellar astrophysics, particularly for irregular variables and quasi-periodic objects such as ultracool brown dwarfs.  Cloud structures forming in the atmospheres of these objects are thought to cause variability over periods of a few hours as the structures rotate into and out of view.  Contemporaneous NIR and optical timeseries observations probe different levels and cloud compositions in these rapidly evolving structures \cite{Vos2019, Tan2021}, but relatively few objects have been observed at NIR and optical wavelengths, due to their intrinsic faintness \cite{Miles2017}.  
Table~\ref{tab:science_cases} gives a concise summary of these scientific opportunities. 

\section{Metrics}
\begin{figure}
\begin{centering}
\includegraphics[width=1.0\textwidth]{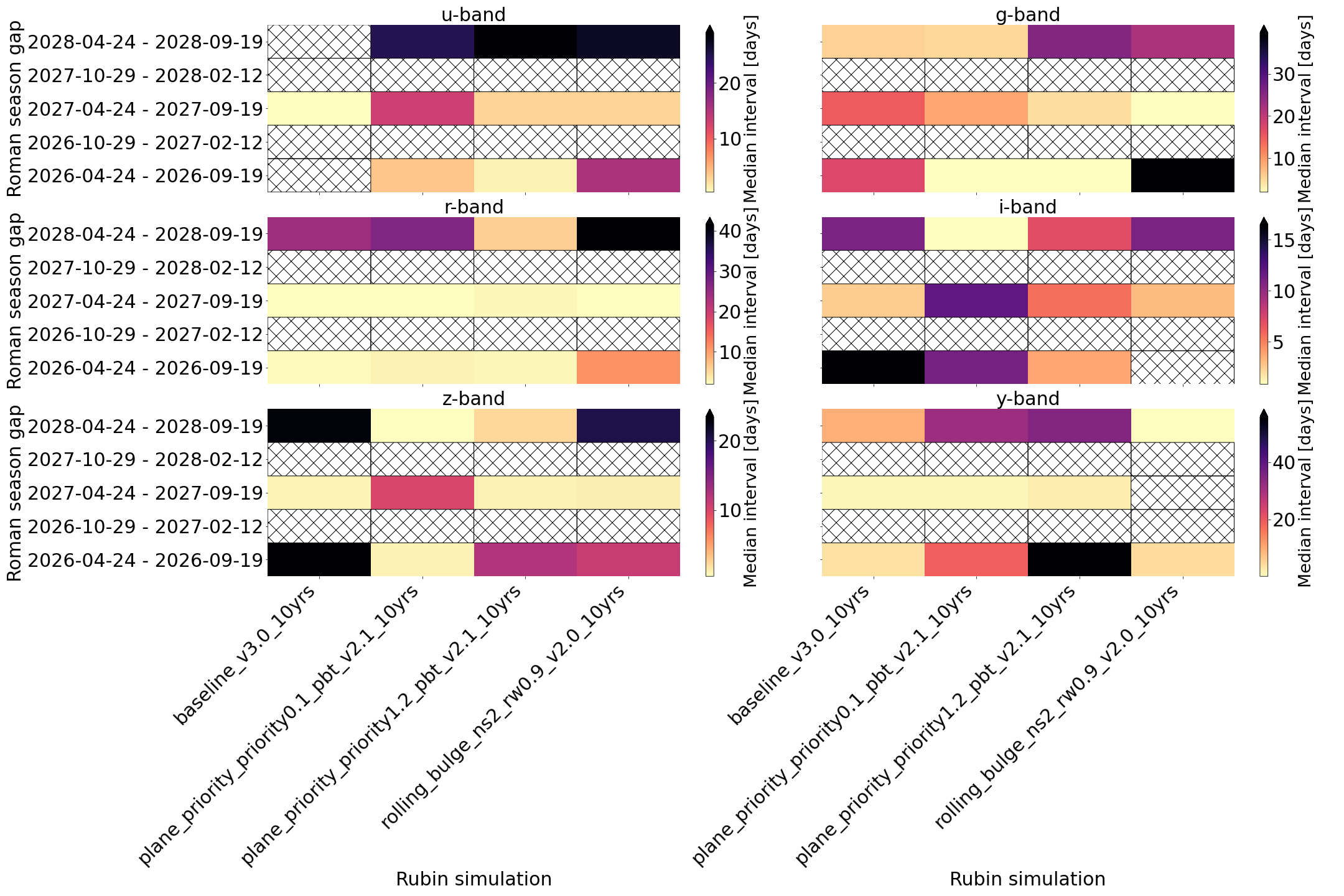}
\caption{Preliminary results from our proposed metric, which evaluates the interval between sequential Rubin observations obtained during the inter-season gaps in the Roman Bulge survey.  Cross-hatching indicates where no observations were obtained. The simulations along the x-axis represent just a few of hundreds of realizations of the Rubin survey strategy under examination. The dates used for the Roman observing seasons are purely nominal at this stage, but reflect our current understanding of the spacecraft's visibility constraints in the direction of the Bulge.
\label{fig:roman_rubin_metric}}
\end{centering}
\end{figure}

We have developed metric software to evaluate whether Rubin visits to the RGBTDS field occur at intervals that complement the RGBTDS seasons.  Our code is designed to be compatible with the Rubin Metric Analysis Framework \cite{MAF} and can be found in our open-source Github repository \cite{RomanRubinMetricRepo}.  This builds on existing MAF metrics ``intervals\_between\_obs\_metric'' and \\``num\_obs\_in\_survey\_time\_overlap\_metric'' by S.~Khakpash. The exact dates of the RGBTDS seasons are to be finalized and can be configured within our metric.  For the time being, the dates used are based on Roman's pointing constraints, allowing it to observe the Bulge.   The metric examines the timestamps of Rubin observations, which it can load from any one of the many simulations of alternative observing strategies explored by the Rubin Observatory.  The code then selects those Rubin observations of the RGBTDS field which occur within the gaps between the RGBTDS observing seasons, selecting only observations which reach a minimum limiting magnitude.  The median interval between sequential Rubin Observations in the inter-season gap (medianed over all seasons), gives a simple numerical value by which different strategies can be compared.  Figure~\ref{fig:roman_rubin_metric} presents the results of this metric comparing just a few of hundreds of Rubin survey strategy simulations.  We recommend that the metric be optimized to a value of 1\,day or less by coordinating the RGBTDS seasons and the LSST observing strategy. 
  
\section{Synergies with other survey facilities}
Although we have focused on synergies with the Rubin Observatory in this White Paper, it will not be the only major survey facility with complementary capabilities, and we recommend a broader review be undertaken to evaluate the benefits of joint data analyses.  For example, the Gaia source catalog is limited to relatively bright stars within the RGBTDS field, as the crowding in this region leads to excessive demands for onboard computations.  Roman photometry and astrometry will extend our view of the Bulge and Bar to regions on the far side of the Galaxy (Fig.~\ref{fig:simgalbulge}).  This promises to offer a goldmine of ages and distance measurements from variable star lightcurves that will be valuable for galactic archaeology and dynamics.   It will also help us to better assess extinction and other aspects of Gaia's selection function, such as crowding. 

\section{Conclusions}
The Roman GBTDS will be groundbreaking not only for exoplanetary science but for a wide range of stellar astrophysics.  That the survey will be in operation at the same time as other wide-field surveys of similar limiting magnitude, sky area, and complementary time cadence and wavelength coverage offers us unique scientific opportunities.  In order to maximize the science return from both Roman GBTDS and the LSST, we recommend:
\begin{enumerate}
    \item That the RGBTDS seasons be scheduled in coordination with the Rubin Observatory's LSST, such that Rubin provides, at minimum, daily observations in at least 3 optical passbands during the intra-season gaps. These data will provide alerts of microlensing anomalies that would otherwise be missed, as well as tighter constraints on the microlensing parallax, and hence the mass of the lensing objects, than Roman alone can achieve.  The same dataset will deliver a wealth of additional astrophysics, helping us to characterize RR~Lyrae in the Bulge, etc.  We further recommend that opportunities for contemporaneous observations be explored.  This will require close coordination between the groups responsible for the survey strategy for the Roman Mission and Rubin Observatory.  
    \item That a broader study of the scientific potential of coordinated scheduling and data analysis from major surveys should be funded.  Due to space constraints we have focused on science from RGBTDS, but the combination of NIR and Rubin's optical data will also be highly beneficial for the High Latitude Survey, for example, in the measurement of photometric redshifts of galaxies, characterizing supernovae lightcurves and probing the edges of the Milky Way halo.  Furthermore, there will be other complementary surveys operating within this timeframe, such as ULTRASAT \cite{ultrasat}, that can further extend the wavelength coverage.  We advocate for close coordination of the survey observing strategies, data handling and metrics of the next generation of Great Observatories and existing catalogs such as those from Gaia.
    \item That metrics be developed to evaluate how changes to one survey's strategy impact the science return from another {\em as a joint effort between Roman, Rubin, and other major surveys.}  
\end{enumerate}
We note that it would be valuable to have a common framework for writing and running survey strategy simulations and metrics, rather than develop separate code bases.  The Metric Analysis Framework \cite{MAF} is an example of a project-supported code base that has successfully integrated metric code contributed from the wider community.  

We emphasize that we do not request any changes to the current design of the Roman GBTDS in terms of changing the footprint or filter selection within a specific season.  Rather, we note the exciting potential benefits of simply coordinating the scheduling of the existing survey design with other facilities. 

\subsection{Synergies with other Roman Core Community Survey White Papers}
These recommendations support the aims of the {\em The Roman Galactic Exoplanet Survey (RGES)} White Paper by Gaudi et al., and are complementary to those of a number of other community White Papers, in particular those of Lam et al {\em ``Characterizing the Galactic population of isolated
black holes''}, and Terry et al. {\em ``A field at the Galactic Center''}.  In support of the latter paper, we note that the Rubin FOV is large enough to include both the current Roman GBTDS footprint and the proposed field at the Galactic Center, in a single pointing.  We also note that Bechtol et al. {\em Coordinating Roman and Rubin for Cosmic Probes of Dark Matter with Resolved Stellar Populations} discuss the benefits of coordination between Roman and Rubin for the High Latitude survey.

\subsection*{Appendix A: Rubin Observatory's Legacy Survey of Space and Time (LSST)}
\label{sec:rubin}
\begin{figure}
\begin{centering}
\includegraphics[width=0.8\textwidth]{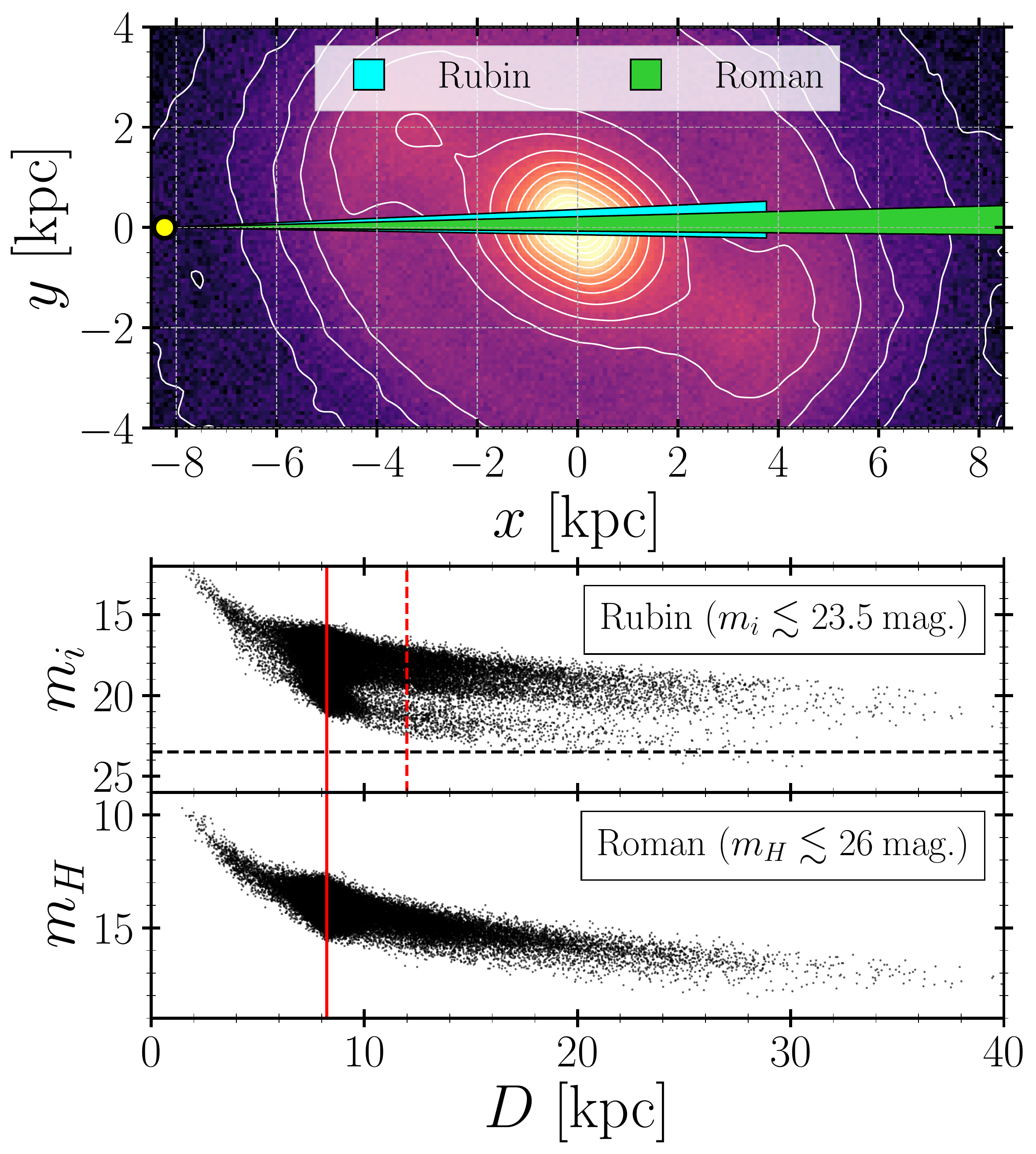} 
\caption{(Top) The face-on density distribution of a Milky Way-like barred galaxy from the Auriga Project \cite{Grand17}.  The bar major axis is angled $30^\circ$ from the $x$-axis. White contours follow lines of constant log density.  The yellow circle shows the location of the approximate Solar position within the model at $X=-8.232\>{\rm kpc}$ \cite{GRAVITY+19, GRAVITY+21}.  The cyan and green wedges represent the widths and approximate depths of the Roman GBTDS footprint and a Rubin pointing, respectively (see Fig.~\ref{fig:bulge_survey_footprint}).  The depth of the Rubin field is limited to a heliocentric distance of $D\lesssim12\>{\rm kpc}$ from the Solar position. (Bottom) Assuming each stellar particle of the model represents a red clump star (RC), we assign absolute magnitudes of $M_i({\rm RC})=0.37\pm0.30$ and $M_H({\rm RC})=-1.40\pm0.30$ \cite{Plevne2020} convolved with a Gaussian kernel (here the $H$-band serves as a proxy for Roman's F146 band).  Then, by converting the positions of stellar particles to heliocentric coordinates, we estimate extinction values using the \texttt{combined19} dust map of the python package \texttt{mwdust} \cite{Bovy16}, allowing us to calculate apparent magnitudes of these mock RC stars.  In the two panels, we present the apparent magnitudes in the $i$- and $H$-bands as a function of $D$, with the red vertical line denoting the Galactic Center and the red vertical dashed line in the top panel showing the estimated limiting depth of the Rubin field.  Rubin $i$-band observations are expected to be limited to $m_i\lesssim23.5\>{\rm mag.}$ (horizontal dashed line) or $D\lesssim12\>{\rm kpc}$, reaching just beyond the Galactic Center.  We predict that Roman's magnitude limit will allow it to probe far beyond the Galactic Center.
\label{fig:simgalbulge}}
\end{centering}
\end{figure}

\begin{figure}
\begin{centering}
\begin{tabular}{cc}
\includegraphics[width=0.45\textwidth]{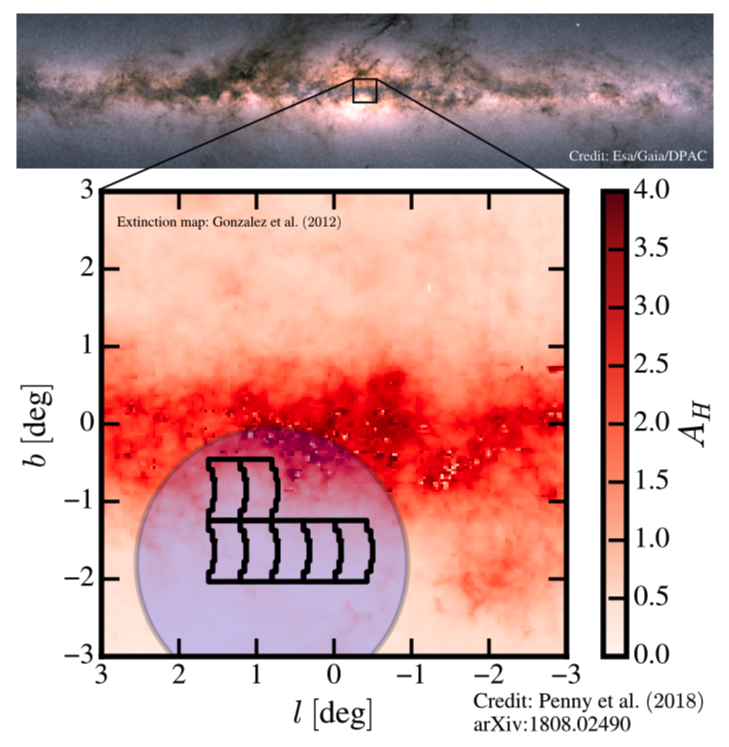} 
\includegraphics[width=0.45\textwidth]{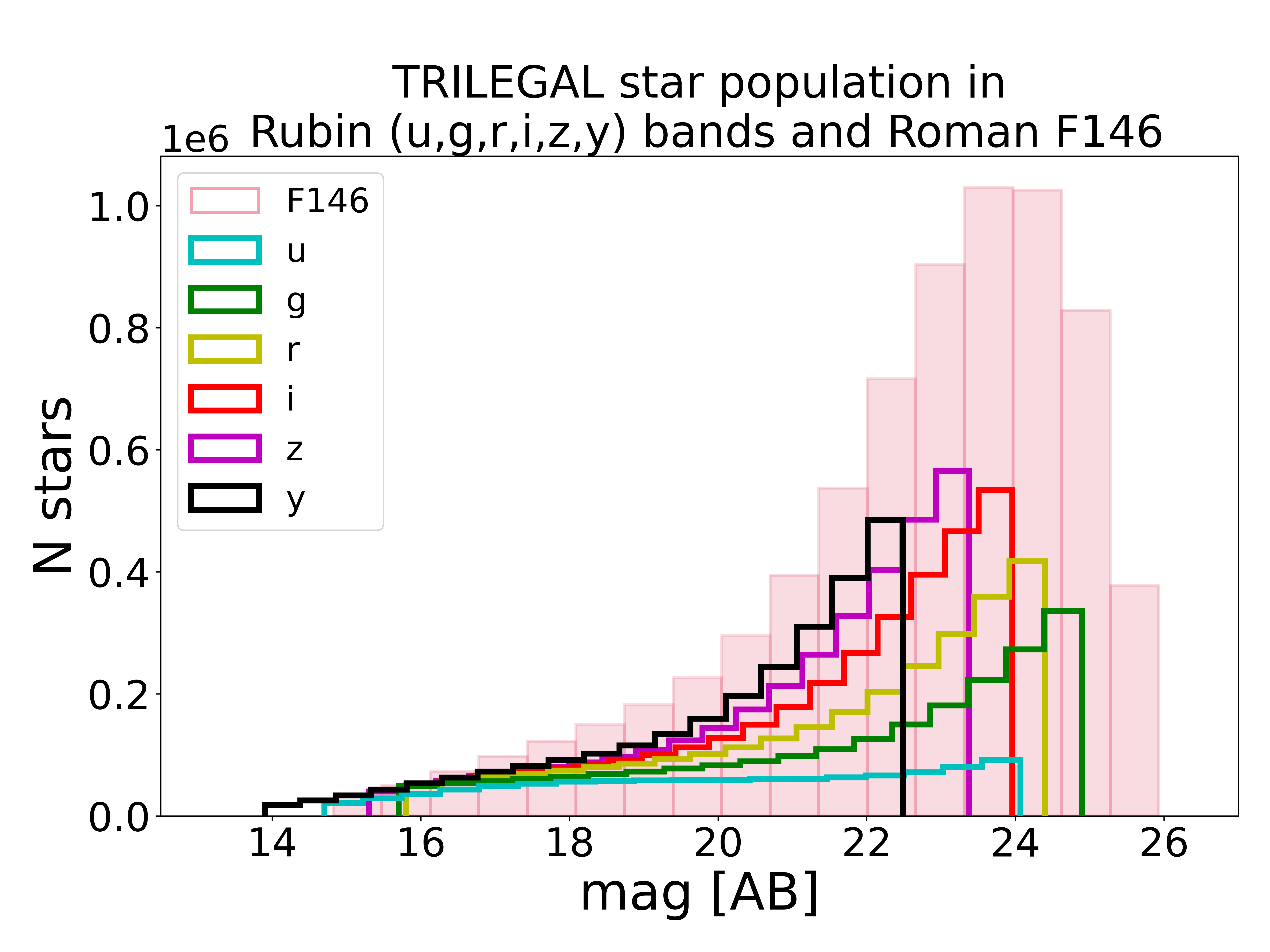} 
\end{tabular}
\caption{(Left) Comparison of the Roman Galactic Bulge Time Domain Survey field (black mosaic outline, 1.53x1.5\,deg) with the field of view of a proposed LSST Deep Drilling Field (blue circle, 3.5\,deg diameter) \cite{Street2018}. Note that the LSST survey strategy is still being refined, and could be adjusted to include a Roman field at the Galactic Center. (Right) Histogram of stellar apparent magnitudes for the Bulge survey field based on the TRILEGAL galactic model, showing that Roman and Rubin will observe complementary stellar samples. \label{fig:bulge_survey_footprint}}
\end{centering}
\end{figure}

With a 9.6\,sq.deg. field of view, 8.4\,m aperture and spatial resolution of 0.2"/pix, Rubin Observatory can deliver optical (SDSS-$u,g,r,i,z,y$) imaging that is highly complementary to that of Roman in the NIR.  Rubin's signature survey, LSST, is expected to begin in early 2025 and continue for 10\,yrs. The details of Rubin's survey strategy are currently being refined \cite{PSTN55}, but the most recent baseline now includes long-term monitoring of a large region of the central Bulge, fully including the RGBTDS survey footprint and operating contemporaneously (Fig.~\ref{fig:bulge_survey_footprint}).  Rubin's limiting magnitude in single exposures of the crowded Bulge fields is expected to reach (u: 24.07, g: 24.90, r: 24.40, i: 23.96, z: 23.38, y: 22.49)\,mag \cite{suberlak2018} (it will probe deeper in high-latitude fields).  Our simulations indicate that Rubin will be able to monitor 47\% of stars detected by Roman (Fig.~\ref{fig:bulge_survey_footprint}) and that Roman will probe tens of kiloparsecs deeper than Rubin (Fig.~\ref{fig:simgalbulge}).  With careful coordination between Roman and Rubin, LSST could deliver regular, long-baseline monitoring of the RGBTDS field that would fill the Roman season gaps. This would provide more precise measurements of the microlensing parallax (both due to Earth's annual motion and the satellite parallax due to the separation of the observatories) and provide real-time alerts of anomalous features.  The optical$+$NIR data will characterize the Spectral Energy Distribution of the microlensing source stars, a vital step in the estimation of the source star's angular radius, which allows the lens mass to be determined.  It will also identify the faintest source stars for future follow-up by Adaptive Optics imaging, once the lens and source stars have separated.  

We note that a number of key elements of Rubin's observing strategy in the Galactic Plane remain to be decided, in particular, whether a `rolling cadence' strategy would be beneficial.  A rolling cadence divides the sky into different spatial regions.  Higher cadence observations can then be performed for one region while the other(s) is observed at lower cadence. In subsequent years the regions are alternated, so that the entire survey footprint eventually receives the same number of visits.  Through close coordination, a Rubin rolling cadence could be applied to a small region, including the RGBTDS footprint, with the high-cadence ($\sim$1--2 visits/day) phase timed to occur during the inter-season gaps in RGBTDS.   This strategy was proposed as a Rubin survey strategy White Paper in \cite{Street2018}.  

\vspace{.6in}

\vspace{.6in}

\vspace{.6in}

\vspace{.6in}

\vspace{.6in}

% Original shared document:
% https://docs.google.com/document/d/1xsuHtiTkVBLsPc7XyEiss2fWZ-FLaopdPoEp1iZZxvw/edit
\begin{table}[!ht]
\begin{tabular}{ |c|c|c|c|c|}
\hline
Science case & Synergy & Roman alone & Rubin Synergy & Sci. refs. \\
\hline
\hline
\multirow{5}{3cm}{Characterizing mid-to-long duration microlensing events} & \hlc[WildStrawberry]{ Gap fill } & \multirow{5}{4cm}{Black holes, neutron stars, some planetary host stars will have large gaps in lightcurve} & \multirow{5}{3.5cm}{* Improved parallax and mass determinations \\ * Source star characterization} & \multirow{5}{1.5cm}{\cite{Lam:2022, Sahu:2022, Lam2020, Wyrzykowski:2016}} \\ 
& \hlc[YellowOrange]{ O + IR } & & & \\ 
& \hlc[Yellow]{ Baseline } & & & \\ 
& \hlc[Orchid]{ Overlap } & & & \\ 
& & & & \\ 
\hline
\multirow{6}{3cm}{Characterizing variable stars (e.g. eclipsing binaries, pulsators long period variables, Miras)} & \hlc[WildStrawberry]{ Gap fill } & \multirow{6}{4cm}{Only red optical$+$IR colors, large gaps in lightcurves} & \multirow{6}{3.5cm}{* More accurate classifications \\ * Period-color- luminosity relations \\ * Photometric metallicities} & \multirow{6}{1.5cm}{\cite{Nikzat:2022,Soszynski:2011,Soszynski:2013,Soszynski:2016,Soszynski:2017}} \\ 
& \hlc[YellowOrange]{ O + IR } & & & \\ 
& \hlc[Yellow]{ Baseline } & & & \\ 
& & & & \\ 
& & & & \\ 
& & & & \\ 
\hline
\multirow{3}{3cm}{Brown dwarf variability} & \hlc[YellowOrange]{ O + IR } & \multirow{3}{4cm}{Only red optical $+$ IR variability} & \multirow{3}{3.5cm}{* Cloud structures \\ * Atmospheric circulation} & \multirow{3}{1.5cm}{\cite{Vos2019,Tan2021}} \\ 
& \hlc[Orchid]{ Overlap } & & & \\ 
& & & & \\ 
\hline
\multirow{3}{3cm}{Stellar populations, exoplanet transit host stars} & \hlc[YellowOrange]{ O + IR } & \multirow{3}{4cm}{Limited host star metallicities} & \multirow{3}{3.5cm}{* Stellar metallicities \\ * More accurate stellar properties } & \multirow{3}{1.5cm}{\cite{Brown:2009,Hejazi2015}} \\ 
& \hlc[YellowGreen]{ Footprint } & & & \\
& & & & \\

\hline
\multirow{6}{3cm}{Interstellar extinction and Galactic structure} & \hlc[YellowOrange]{ O + IR } & \multirow{6}{4cm}{Only red+IR extinction, deep and precise astrometric solutions} & \multirow{6}{3.5cm}{* O + IR extinction comparison \\ * Characterize standard candles through Milky Way Bulge/Bar} & \multirow{6}{1.5cm}{\cite{Clarkson2018, GoughKelly22,Nataf:2013,Nataf:2016}} \\ 
& \hlc[YellowGreen]{ Footprint } & & & \\
& & & & \\
& & & & \\
& & & & \\
& & & & \\
\hline
\end{tabular}
\caption{Benefits of Roman + Rubin coordination.
We highlight several synergies: Rubin filling in Roman gaps (\hlc[WildStrawberry]{ Gap fill }), Rubin optical complementing Roman IR (\hlc[YellowOrange]{ O + IR }), Rubin’s temporal baseline and start before Roman (\hlc[Yellow]{ Baseline }), Rubin’s larger footprint than Roman (\hlc[YellowGreen]{ Footprint }), Roman + Rubin overlapping/contemporaneous multi-wavelength observations (\hlc[Orchid]{ Overlap }).
\label{tab:science_cases}}
\end{table}

\printbibliography

\end{document}